\documentstyle[twocolumn,amssymb,aps,psfig]{revtex}
\tighten
\begin{document}
\draft
\title{\bf Disorder effects in diluted ferromagnetic semiconductors.
}
\author{G.~Bouzerar$^{a}$, J.~Kudrnovsk\'y$^{a,b}$ and P.~Bruno$^{a}$}
\address{
$^{a}$Max-Planck-Institut f\"ur Mikrostrukturphysik, Weinberg 2,
D-06120 Halle, Germany \\
$^{b}$Institute of Physics AS CR, Na Slovance 2, CZ-182 21
Prague, Czech Republic
 }
\address{~
\parbox{14cm}{\rm
\medskip
Carrier induced ferromagnetism in diluted III-V semi-conductor is
analyzed within a two step approach. First, within a single site
CPA formalism, we calculate the element resolved averaged Green's
function of the itinerant carrier. Then using a generalized RKKY
formula we evaluate the Mn-Mn long-range exchange integrals and
the Curie temperature as a function of the exchange parameter,
magnetic impurity concentration and carrier density. The effect
of the disorder (impurity scattering) appears to play a crucial
role. The standard RKKY calculation (no scattering processes),
strongly underestimate the Curie temperature and is inappropriate
to describe magnetism in diluted magnetic semi-conductors. It is
also shown that an antiferromagnetic exchange favors higher Curie
temperature.
}}
\maketitle

\narrowtext After the recent discovery by Ohno et al. that by
doping GaAs \cite{Ohno} with only $5\%$ of magnetic impurities
Mn$^{2+}$, the Curie temperature $T_{C}$ could already exceed 100
K and because of all the possible technological applications, the
interest for the III-V diluted magnetic semiconductors (DMS) has
increased considerably. In spite of the apparent success of
different methods (mean field, first-principle, random-phase
approximation (RPA)) where disorder is either neglected or
treated at the lowest order to reproduce the Curie
temperature\cite{Dietl,Jungwirth,Koenig1,Sanvito,Akai,Shirai,Bouz1},
there is still some shadow region concerning the effect of
disorder on magnetism. Indeed only few works, mainly based on
numerical simulations, are including the effect of positional
disorder \cite{Bhatt,Schlim} within a mean field treatment
(multi-scattering effects are not included). Recently, in order
to provide a simultaneous and self-consistent treatment of the
itinerant carrier and magnetic impurity an approach based on the
equation of motion method was proposed \cite{Bouz1}. However, as
a consequence of the RPA decoupling the self-energy of the
itinerant carriers Green's function (GF) is reduced to the lowest
order term $\Sigma_{\sigma} = \frac{z_{\sigma}}{2} J_{pd} c \langle \hat{\bf S}^{z} \rangle$ (where $z_{\sigma} =\pm 1$ and c is Mn$^{2+}$ concentration). 
Because of the difficulty to include within this
formalism, higher order scattering terms, we follow in this paper
a slightly different approach and focus first on the effect of
the disorder on the itinerant carriers. First, we calculate the
itinerant carrier GF by treating the effect of
disorder in the full coherent-potential approximation (CPA)
approximation, which means that all single site multi-scattering
processes are properly included. In the second step we calculate
the exchange integrals between spin impurities using the
projected GF on the Mn sites. The difficulty is to
perform properly the averaged T-matrix calculation since the
holes/electrons scattering depends on the impurity spin operator.
For that purpose we follow the procedure described in
\cite{Taka,Kubo}. It should be added that in this work, spin
impurities are treated fully quantum mechanically.

We consider the following minimal Hamiltonian which is the good
starting point to study DMS,

\begin{equation}
H=\sum_{ij, \sigma} t_{ij}c_{i\sigma}^{\dagger}c_{j\sigma}+
\sum_{i}J_{i} {\bf S}_{i}\cdot {\bf s}_{i}
\end{equation}
$t_{ij}=t$ for $i$ and $j$ nearest neighbors and zero otherwise.
In the exchange between localized impurities spin and itinerant
electron gas $J_{i}$ is a random variable: $J_i=J$ ($J \ge 0$
means antiferromagnetic coupling) if the site $i$ is occupied by a
magnetic impurity, and zero otherwise. ${\bf S}_{i}$ is the
magnetic impurity spin operator at site $i$ and ${\bf
s}_{i}=c_{i\alpha}^{\dagger} (1/2\mbox{\boldmath$\sigma$}_{\alpha
\beta})c_{i\beta}$ is the spin operator at site $i$ of the
itinerant electron gas.

The T-matrix associated to the multi-scattering of a single
magnetic impurity (at site $m$) embedded in the effective medium
is,

\begin{equation}
\hat{\bf t}_{m}=\hat{\bf V}_{m} ({\bf 1}-\hat{\bf \bar{G}}\hat{\bf V}_{m})^{-1}
\end{equation}
where,
\begin{eqnarray}
\hat{\bf V}^{\text{Mn}}_{m}=\left(
\begin{array}{cc}
 \frac{1}{2}J {\bf S}_{m}^{z}-\Sigma_{\uparrow} & \frac{1}{2}J {\bf S}_{m}^{-} \\
 \frac{1}{2}J {\bf S}_{m}^{+} & -\frac{1}{2} J {\bf S}_{m}^{z}-\Sigma_{\downarrow}
\end{array}
\right)
\end{eqnarray}
Respectively, for Ga at site $m$, $\hat{\bf V}^{\text{Ga}}_{m}$
is obtained by taking $J=0$ in the previous equation.

The $2 \times 2$ averaged Green's function matrix $\hat{{\bf \bar{G}}}$ is,
\begin{eqnarray}
\hat{{\bf \bar{G}}}=\left(
\begin{array}{cc}
\hat{\bf \bar{G}}_{\uparrow} & 0 \\ 0 & \hat{\bf \bar{G}}_{\downarrow}
\end{array}
\right)
\end{eqnarray}
with $\hat{\bf \bar{G}}_{\sigma}=(\omega{\bf I} -\hat{\bf {\bar{K}}}_{\sigma})^{-1}$ where, $\hat{\bf \bar{K}}_{\sigma}= \sum_{\bf k}
(\epsilon_{\bf k} - \Sigma_{\sigma}) c^{\dagger}_{{\bf k}\sigma}c_{{\bf k}\sigma}$. 

In the following, we omit the site index $m$.
The self-energy $\Sigma _{\sigma}$
is obtained by solving the coupled self-consistent equations.

\begin{eqnarray}
\langle t_{\sigma \sigma} \rangle_{\text{dis},T} = (1-c)\langle
t^{\text{Ga}}_{\sigma \sigma} \rangle_{T}+ c\langle
t^{\text{Mn}}_{\sigma \sigma} \rangle_{T}=0
\end{eqnarray}
$\langle... \rangle_{\text{dis},T}$ denotes configuration average
and thermal average at temperature $T$ for the spin operator, $c$
is the concentration of Mn impurities.

After lengthy calculations we get,
\begin{eqnarray}
\Sigma_{\sigma}= \frac{\Large \langle \sum_{\lambda}
x_{\lambda}\hat{\bf c}^{\lambda}_{\sigma}(\hat{\bf
d}^{\lambda}_{\sigma})^{-1} \Large \rangle_{T}} {\Large \langle
\sum_{\lambda} x_{\lambda}(\hat{\bf d}^{\lambda}_{\sigma})^{-1}
\Large \rangle_{T}} \label{self}
\end{eqnarray}
The sum runs over all constituents, in our binary system
$x_{\text{Mn}}=c$ (resp. $x_{\text{Ga}}=1-c$).
\begin{eqnarray}
{\hat{\bf c}}^{\text{Mn}}_{\sigma}&=&
z_{\sigma}\frac{J}{2}\hat{\bf S}^{z}+\large( \frac{J^{2}}{4}
[S(S+1)-(\hat{\bf S}^{z})^2 -z_{\sigma}\hat{\bf S}^{z}]  \nonumber \\
&&+ [z_{\sigma}\frac{J}{2}\hat{\bf
S}^{z}+\Sigma_{\sigma}][\frac{J}{2}(z_{-\sigma}\hat{\bf S}^{z}-1
)-\Sigma_{-\sigma}]   \Large) G_{-\sigma}
\end{eqnarray}
and,
\begin{eqnarray}
{\hat{\bf d}}^{\text{Mn}}_{\sigma}&=&1-G_{-\sigma}[\frac{J}{2}(z_{-\sigma}\hat{\bf S}^{z}-1)-\Sigma_{-\sigma}] \nonumber  \\
&&- G_{\sigma}[{\hat{\bf
c}}^{\text{Mn}}_{\sigma}-z_{\sigma}\frac{J}{2}\hat{\bf S}^{z}]
\end{eqnarray}
with $z_{\sigma}=1$ (resp.$-1$) for spin $\uparrow$ (resp.$\downarrow$). 
$\bar{G}_{\sigma} = \frac{1}{N} \sum_{\bf q} \bar{G}_{\sigma}({\bf
q},\omega)$ where ${\bar G}_{\sigma}({\bf q},\omega)=\left({\omega
-\epsilon_{\bf q} -\Sigma_{\sigma}(\omega)}\right)^{-1}$ denotes
the averaged GF. Similarly, one gets ${\hat{\bf
c}}^{\text{Ga}}_{\sigma}$ and ${\hat{\bf d}}^{\text{Ga}}_{\sigma}$
after setting $J=0$ in the previous equations.

Note that since impurity spins are treated {\it quantum
mechanically}, the thermal averaged quantities are evaluated
using the following decomposition, $\Large \langle \hat{\bf O}
(\hat{\bf S}^{z})\Large \rangle_{T} = \sum_{i=0}^{2S} a_{i} \Large
\langle (\hat{\bf S}^{z})^{i}\Large \rangle_{T}$ where $\hat{\bf
O}$ denotes a general operator which depends in a non trivial
manner on $(\hat{\bf S}^{z})^{i}$. Additionally, as it was shown
by Callen-Strikman, $\Large \langle (\hat{\bf S}^{z})^{i}\Large
\rangle_{T}$, and hence $\Large \langle \hat{\bf O} (\hat{\bf
S}^{z})\Large \rangle_{T}$, are universal functions of $\Large
\langle \hat{\bf S}^{z}\Large \rangle_{T}$ only \cite{Callen}.
After solving the coupled set of equations (Eq.~(\ref{self}) with
$\sigma =\pm 1$) one gets the total averaged GF of
the itinerant carriers.

The next step consists in calculating the long-range exchange
integrals $J^{\text{eff}}_{ij}$ between magnetic impurities for
the effective Heisenberg Hamiltonian,
\begin{eqnarray}
H^{\text{Heis}}= \frac{1}{2} \sum_{i\ne j}J^{\text{eff}}_{ij}
{{\bf S}_{i}}\cdot {{\bf S}_{j}}
\end{eqnarray}

The exchange integrals between two impurities separated by a
distance ${\bf R}$ is given by the generalized RKKY formula,
\begin{eqnarray}
J^{\text{eff}}({\bf R})= -\frac{1}{2} J^{2} [-\frac{1}{\pi}
\rm{Im} \chi({\bf R})]
\end{eqnarray}
where the susceptibility is,

\begin{equation}
\chi({\bf R})=\!\sum_{\bf k,q}\!\! \int\!\! d\omega f(\omega)
\bar{G}^{\text{Mn}}_{\uparrow}({\bf
k},\omega)\bar{G}^{\text{Mn}}_{\downarrow}({\bf k+q},\omega)
e^{i{\bf q}\cdot{\bf R}}
\label{chi}
\end{equation}
The chemical potential $\mu$ entering the Fermi-Dirac function
$f(\omega)$ is determined at each temperature by fixing the
itinerant carrier density. Note that the exchange integrals are
$T$-dependent through the averaged GF. When
replacing $\bar{G}_{\sigma}$ by the free particle GF the exchange integrals reduce to the standard RKKY.
Additionally, it is important to stress that to calculate
$J^{\text{eff}}_{ij}$, one has to take into account that both site
$i$ and $j$ should be occupied by Mn atom. Thus the non local
GF which enters Eq.\ref{chi}
should be the Mn-resolved GF but not the full averaged one.

To derive the projected GF on Mn sites, we
essentially follow the procedure described in Ref.\cite{Joseph},
which gives,
\begin{eqnarray}
\bar{G}^{\text{Mn}}_{\sigma}({\bf k},\omega)&=&
F_{\sigma}(\omega)(1-F_{\sigma}(\omega)) \bar{G}_{\sigma}(\omega)
\nonumber \\
&& + F^{2}_{\sigma}(\omega)\bar{G}_{\sigma}({\bf k},\omega)
\end{eqnarray}
where $\bar{G}_{\sigma}(\omega)=\frac{1}{N}\sum_{\bf
k}\bar{G}_{\sigma}({\bf k},\omega)$ and
$F_{\sigma}(\omega)=\left({1-\bar{G}_{\sigma}(\omega)
(V_{\text{eff},\sigma}^{\text{Mn}}-\Sigma_{\sigma})}\right)^{-1}$.
The determination of the effective potential
$V_{\text{eff},\sigma}^{\text{Mn}}$ leads to,
\begin{eqnarray}
V_{eff,\sigma}^{Mn}=\Sigma_{\sigma} \frac{1+\Sigma_{\sigma}
\bar{G}_{\sigma}}{c+\Sigma_{\sigma}\bar{G}_{\sigma}}
\end{eqnarray}
As a final step we can evaluate the Curie temperature, by using
mean-field theory for the effective Heisenberg model:
\begin{eqnarray}
k_{B}T_{C} = \frac{2}{3} S(S+1)c \frac{1}{N}\sum_{\bf q} E(\bf q)
\end{eqnarray}
$E({\bf q})$ is the $T$-dependent magnon spectrum:
 $E({\bf q})={\tilde J}^{\text{eff}}({\bf 0}) - {\tilde J}^{\text{eff}}({\bf q})$where ${\tilde J}^{\text{eff}}({\bf q})$ denotes the Fourier transform
of the exchange integrals.

In the following we discuss the numerical results. In Fig.~1 we
have plotted the total density of states (DOS) and the projected
one on Mn site as a function of energy, for different values of
the parameter $J/t$. We observe that in the weak coupling
regime($J/t=0.86$) the total spin-resolved DOS is almost
identical to the unperturbed one although the Mn-projected DOS is
already strongly affected by disorder. By increasing further
$J/t$, we observe at low energy the impurity band formation. Note
that the impurity band splits first at $E \le 0$, and the
position of the peaks in the Mn-DOS are not symmetric with respect
to 0. This can be easily understood by analyzing the atomic limit
($J/t \rightarrow \infty$) which is properly described. In the
paramagnetic phase, we get a peak at
$E_{\text{high}}=+\frac{1}{2}JS$ and another at
$E_{\text{low}}=-\frac{1}{2}J(S+1)$ with respective weights
$\frac{S+1}{2S+1} c$ and $\frac{S}{2S+1} c$ \cite{note1}.

In Fig.~2, the dependence of $T_{C}$ on $\gamma=n_{h}/c$ ($n_{h}$
is the hole concentration) is discussed. At fixed $\gamma$, the
Curie temperature increases significantly with $J/t$ and large
value are reached when approaching the split-band regime. In the
intermediate regime ($J/t \ge 2$), $T_{c}$ appears to be {\it very sensitive} 
to $J/t$, a maximum at $\gamma \approx 0.10$ is
observed before $T_{C}$ decrease and eventually vanishes at
$\gamma_{c}$ which is $J/t$ dependent. These results are
qualitatively comparable to those of ref.\cite{DasSarma}, although we
obtain Curie temperature significantly larger. Additionally, in
comparable regime the maximum of $T_{c}$ in ref.\cite{DasSarma} 
corresponds to half-filled impurity band ($\gamma \approx 0.50$) and $T_{c}$
is symmetric with respect to this point (it vanishes at $\gamma
=1$). Later it will be shown that the sign of $J$ which is irrelevant in
most of the model calculations, plays in fact an important role. As it
will be discussed in the next section, the value of $\gamma_{c}$
for which $T_{C}$ vanishes depends on both the sign and amplitude
of $J/t$.

In Fig.~3, $T_{C}$ as a function of $J/t$ is shown for both
antiferromagnetic and ferromagnetic coupling for different carrier
density. First, the sign of $J/t$ appears to be
relevant. Indeed, $T_{C}$ is strongly asymmetric with respect to
$J/t=0$. In the case of ferromagnetic coupling the maximum of
$T_{C}$ is {\it much smaller} that for antiferromagnetic coupling.
However, as expected, for $|J|/t \le 1$ they are comparable and
reduce to the standard RKKY calculations ($T_{C} \propto J^{2}$).
For both, ferromagnetic and antiferromagnetic coupling, the
position of the maximum depends on the hole concentration.
However, the maximum occurs earlier in the ferromagnetic case.
Note that in the {\it intermediate regime} $1 \le J/t \le 3$,
$T^{\text{RKKY}}_{C}$ is {\it much smaller} than $T_{C}^{\text{CPA}}$:
for $J/t=2$, $T^{\text{CPA}}_{C} \approx 3 ~T_{C}^{\text{RKKY}}$
for $n_{h}=0.015$. Additionally, after the maximum is reached,
$T_{C}$ drops rapidly, and vanishes at a $n_{h}$-dependent
value of $J/t$. We observe that in the split-band regime $J/t \ge
3.5$ {\it no ferromagnetic ordering} is possible. In contrast to other
approaches, our theory appears to be more suitable to describe the
large coupling regime.

Let us discuss briefly, the relevance of our results with
respect to experimental data of Ga$_{c}$Mn$_{1-c}$As. Our model
is based on a one-band model, as done in Ref.\cite{Bouz1}, we fix
$t$ by assuming a hole effective mass $m^{*} =0.5 ~m_{e}$. This leads
to a value $t=0.58$~eV \cite{note2}. We assume that the
$5.3\%$-doped sample (highest $T_{C}=110$~K) contains
$n_{h} \approx 0.2~c$. Using Fig.4, we obtain $J=-1.12$~eV to get the same
Curie temperature\cite{note3}. Surprisingly, although our calculations are
done within a one-band model, this value agrees perfectly with the
estimate $J=-1.1 \pm 0.2$~eV \cite{Okabayashi}.

In Fig.~4 we analyze the dependence of $T_{C}$ on the impurity
concentration $c$, for different values of $\gamma$. For a given
$c$, we observe that, $T_{C}$ is non monotonic with respect to
$\gamma$. However it is clear, that even at large concentration
the low hole density is more favorable to get a high Curie
temperature. More precisely $T_{C}$ is larger when $\gamma \approx
0.1$. For instance, when $c=0.15$ we get $T_{C} \approx 240$~K.
Additionally, we see that when increasing $c$, for sufficiently
large $\gamma$, $T_{C}$ shows a maximum and decreases till it
vanishes. It is expected that $T_{C}$ will first vanish for larger
itinerant carrier density.Indeed, the localization effect is
stronger at higher carrier density.

To conclude, we have presented a theory to study ferromagnetism
in DMS, which consists (i) in treating the itinerant carrier
within the best single site approximation (CPA) and (ii)
performing the susceptibility calculation with the disordered
Green's functions to get the Curie temperature. We have shown
that the role of disorder is important and leads to significantly
higher values of the Curie temperature with respect to a standard
RKKY calculation. Additionally, it is shown that an
antiferromagnetic coupling favors a higher $T_{C}$ in the hole
doped materials. We believe that this approach is more suitable
to analyze ferromagnetism in diluted semiconductors.

One of us (J.K.) acknowledges the financial support provided by the Grant
Agency of the Czech Republic (No.\ 202/00/0122) and the Grant Agency of
the Academy of Sciences of the Czech Republic ((No.\ A1010203).

%

\begin{figure}
\caption[]{ Total DOS (left panels) and Mn projected DOS (right)
as a function of $2 E/W$ (bandwidth $W=12 t$), for majority spin (dashed
curve) and minority spin (full curve) at $T= 0$~K. The value of
$J/t$ are $J/t= 3.45$ (a), $J/t= 2.33$ (b) and $J/t= 0.86$ (c).
The concentration of impurity is $c=5\%$. } \label{fig1}
\end{figure}

\begin{figure}
\caption[]{$T_{C}/t$ as a function of $\gamma=n_{h}/c$ for
different values of $J/t$.
 The concentration of impurity is fixed to $c=5\%$.
}
\label{fig2}
\end{figure}

\begin{figure}
\caption[]{ $T_{C}/t$ as a function of $J/t$ for different carrier
density. The impurity concentration is $c=5\%$. The RKKY
calculation corresponds to the continuous, dashed and dotted
curves and the full CPA treatment to symbols. } \label{fig3}
\end{figure}

\begin{figure}
\caption[]{ $T_{C}$ (in K) as a function of $c$ for different
values of $\gamma$ (see fig.). The parameters are $t=0.58$~eV and
$J=1.12$~eV. (see text)} \label{fig5}
\end{figure}


\begin{references}
\bibitem{note2} In GaAs the volume of the unit cell is $v=a_{0}^{3}/4$ where
$a_{0}=0.565$~nm. For simplicity, our calculations are performed
on a simple cubic lattice, thus the lattice spacing we take is
$a_{1}=(a_{0}^{3}/4)^{1/3}$.
\end{references}
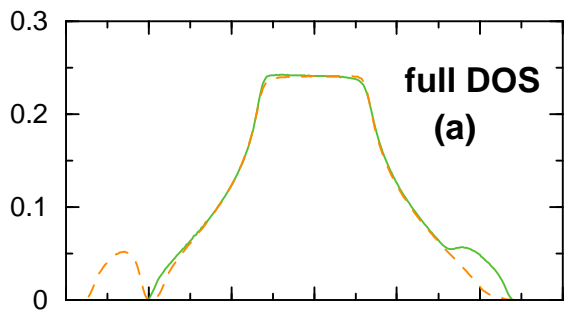
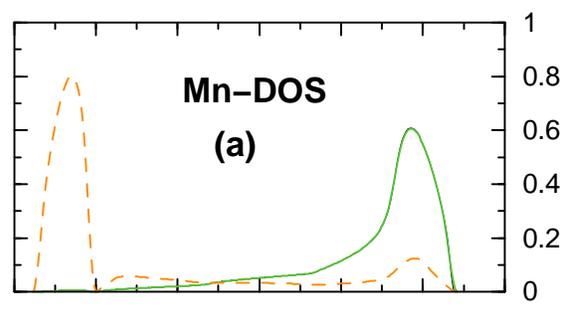
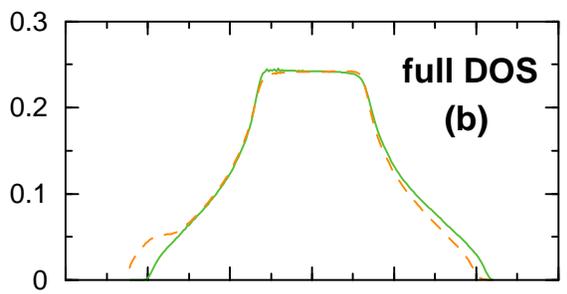
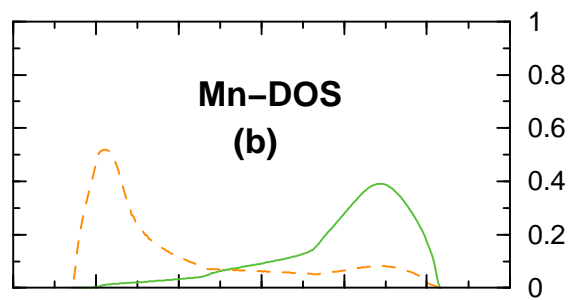
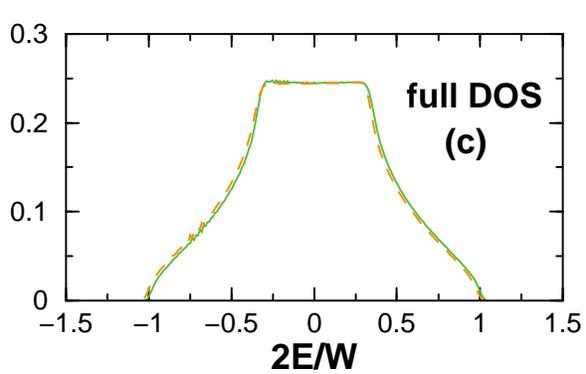
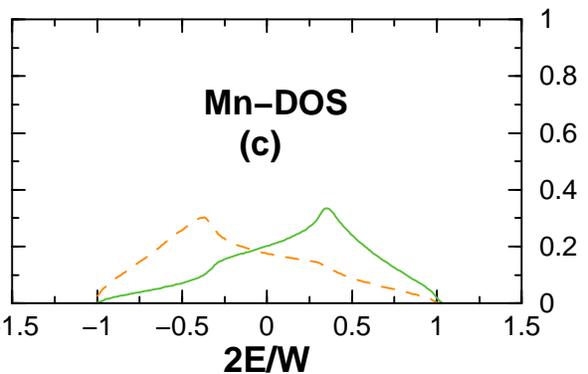
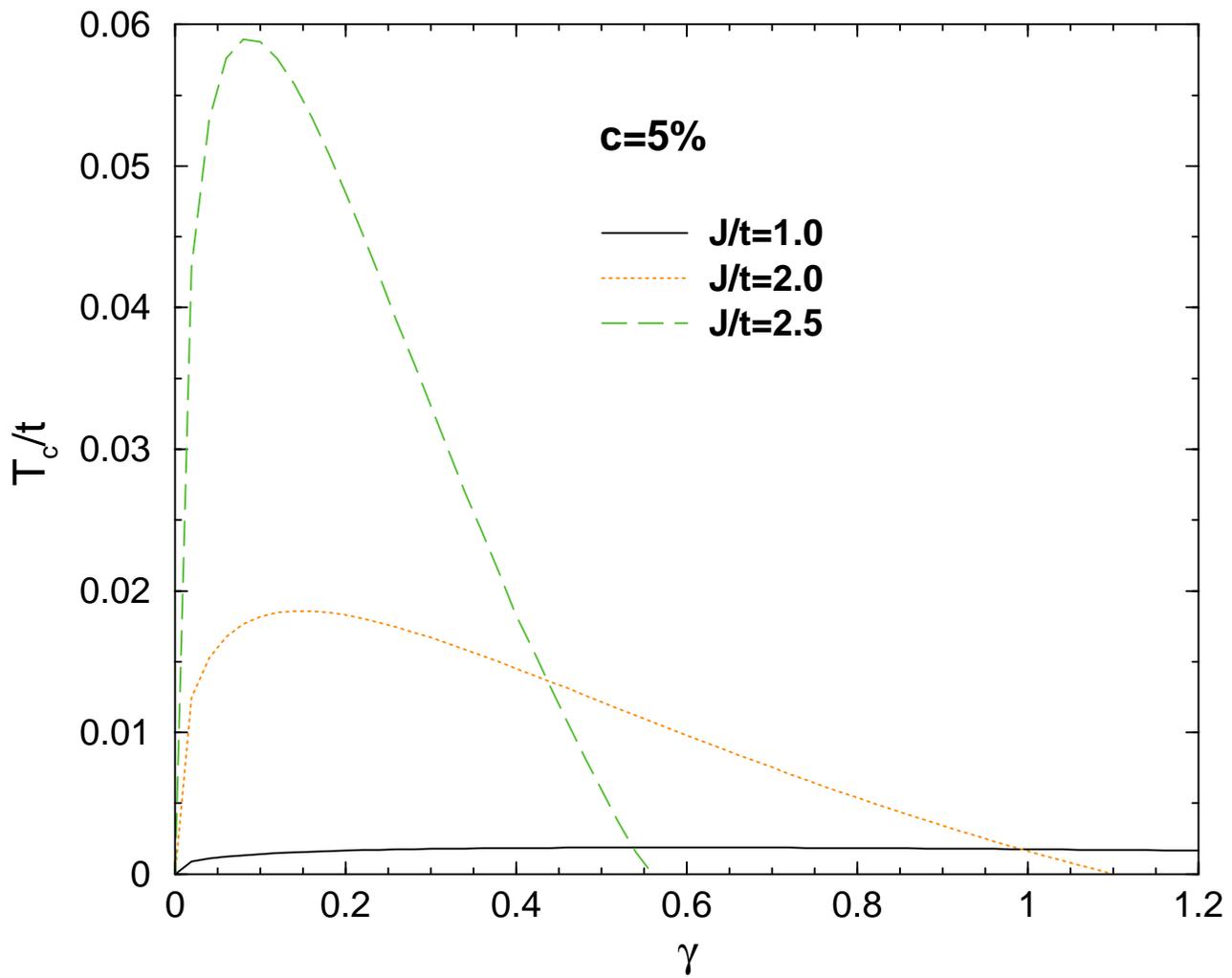
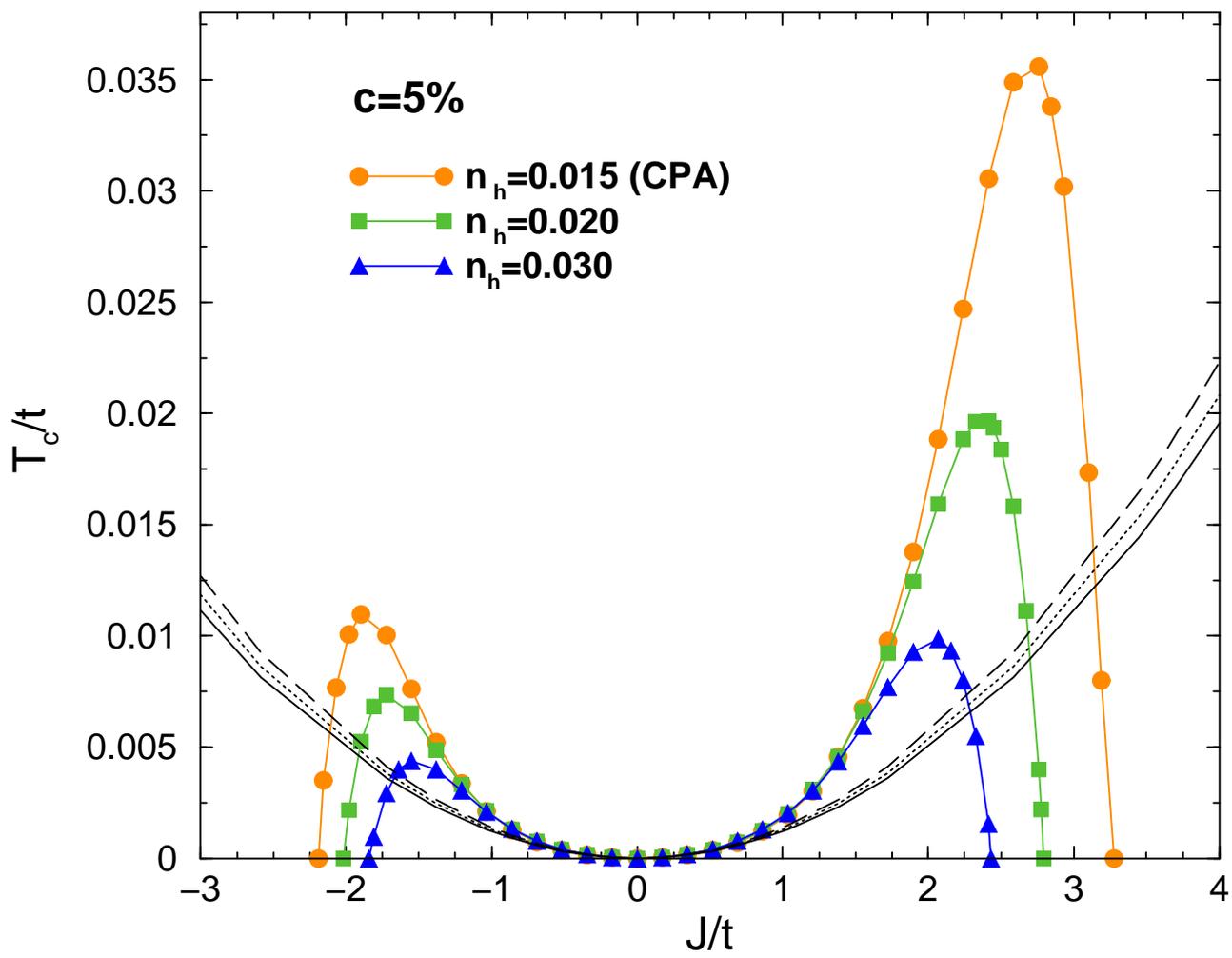
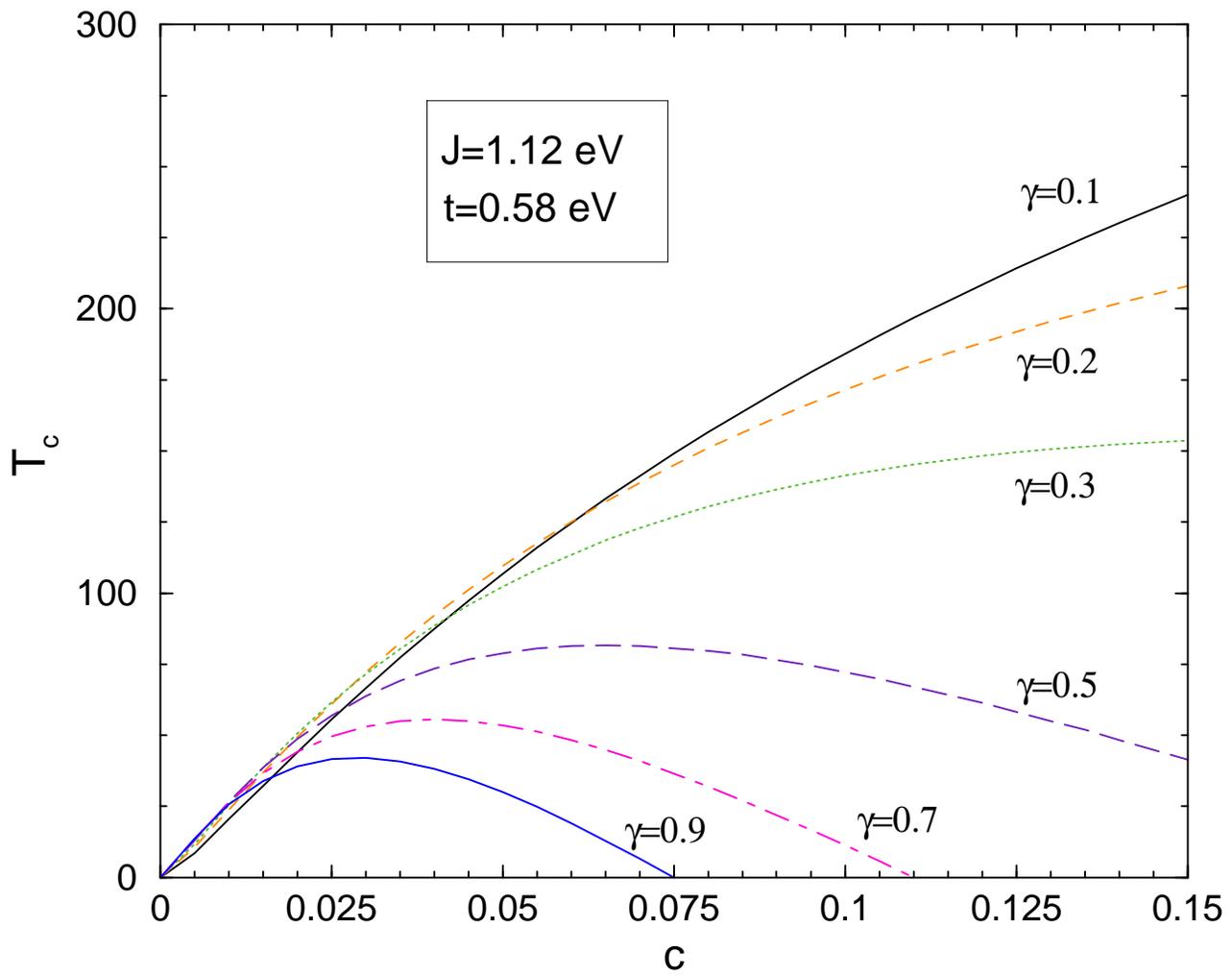
\end{document}